\documentclass[11pt]{article}
\usepackage{moriond,epsfig}

\def\be{\begin{equation}}
\def\ee{\end{equation}}
\def\bea{\begin{eqnarray}}
\def\eea{\end{eqnarray}}

\newcommand{\Eslash}{\mbox{$\rm E \kern-0.6em\slash$}}
\def\etmiss{\mbox{$\rm \Eslash_{T}\!$}}

\newcommand{\etmissscal}{\mbox{$\Eslash_T^{\rm{Scaled}}$ }}
\begin{document}

\vspace*{4cm}
\title{Searching for Higgs Decaying to \\ $H\rightarrow WW^* \rightarrow \mu + \tau_{had}$ and $H\rightarrow WW^* \rightarrow ee$ at $D\O$ } 

\author{ Bj\"orn Penning }

\address{Institute of Physics, Hermann-Herder-Str.3,\\
D-79104 Freiburg, Germany}

\maketitle\abstracts{
A search for the Higgs boson in $H \rightarrow W W^* \rightarrow ee$ and $H \rightarrow W W \rightarrow  \mu \tau_{had}$ decays in $p \overline{p}$ 
collisions at a center-of-mass energy of $\sqrt{s}=1.96$ TeV is presented. The data have been collected by the Run II $D\O$ detector. In order to maximize the sensitivity multivariate techniques such as artificial neural networks  (NN), matrix element methods and likelihoods are used. No excess above the {\it Standard Model} background is observed and limits on the production cross section times  branching ratio $\sigma \times BR(H \to WW^* \to ee)$ for Higgs masses between 115 and 200 GeV are set.
}

\section{Introduction}
Two searches for the Higgs boson decaying to the $WW^*$ final state are presented. The dileptonic decay mode with two electrons in the final state and $H \rightarrow W W \rightarrow \mu + \tau_{had}^{1-prong}$  leading to final states with one muon, a jet originating from a hadronically decaying tau and missing transverse momentum have been studied. The $\mu + \tau_{had}$ analysis has been performed using an integrated luminosity of about $\sim 1$ fb$^{-1}$ of RunII data recorded between 2001 and 2006, known as RunIIa. The $ee$ analysis is performed using data from June 2006 until August 2007, known as RunIIb, corresponding to an integrated luminiosity of $1.2$ fb$^{-1}$. These decay modes in combination with other dileptonic decay modes provide the best sensitivity to a Standard Model (SM) Higgs boson search at the Tevatron at a mass of $m_{\rm H} \sim 160 ~GeV/c^2$ \cite{turcot,tevhiggs,jakobs}. In order to maximize the signal to background separation  multivariate techniques are used. If combined with searches exploiting the $WH$ and $ZH$ associated production, these decay modes increase the sensitivity for the Higgs boson searches.

\section{Event Selection \label{ref:sel}}
The signal is characterized by two leptons, missing transverse momentum ($p_{T}$) and little jet activity. For the $\mu + \tau_{had}$ selection one isolated muon with $p_{T}^{\mu} >$ 12 GeV and one isolated tau with $p_{T}^{\tau} >$ 10 GeV are required. A hadronically decaying tau lepton is characterized by a narrow isolated jet with low track multiplicity. Three tau types are distinguished:
\begin{itemize}
  \item $\tau$-type I: A single track with a calorimeter cluster without any electromagnetic subclusters (1-prong, $\pi$-like).
  \item $\tau$-type II: A single track with a calorimeter cluster and electromagnetic subclusters (1-prong, $\rho$-like).
  \item $\tau$-type III: Two or three tracks with an invariant mass below 1.1 or 1.7 GeV, respectively (3-prong).
\end{itemize}
Due to the large background contamination $\tau$-type III is neglected in the analysis.\\
For the dielectron analysis the leading electron is required to have $p_{T}^{e_1} >$ 20 GeV and the trailing electron to fulfill $p_{T}^{e_2} >$ 15 GeV. Subsequently most of the $QCD$ background is removed by selection requirements on the missing transverse energy $\etmiss$ and the scaled missing transverse energy \etmissscal, defined as $ \Eslash_T^{\rm{Scaled}} = \frac{\etmiss}{\sqrt{\sum_{jets} \sigma^2_{E_T^j||\etmiss}}}$, which is the $\etmiss$ divided by the $\etmiss$ resolution. This quantity is particularly sensitive to events where the missing energy could be a result of mismeasurements of jet energies in the transverse plane. A requirement on the minimal transverse mass, $M_T^{min}=\sqrt{2 \cdot \etmiss \cdot p_T^l \cdot (1-\cos (\Delta \phi))}$,  between one of the leptons and $\etmiss$ reduces further the various background processes. Most of the $\rm Z/\gamma^{*} \rightarrow \ell \ell$ events are rejected by requiring the sum of the momentum of  $p_T^{\mu} + p_T^{\tau} + \etmiss$ to be within given lower and upper boundaries and by requiring the invariant mass to be less than the $Z$ peak. The  $\rm t\bar{t}$ contribution is reduced by requiring low values $H_{T}$ which is defined as the scalar sum of the transverse momenta of all jets in the event. A large fraction of remaining back-to-back  $\rm Z/\gamma^{*} \rightarrow \ell \ell$ is reduced by rejecting events with a wide opening angle between the leptons. Since the signal kinematics change as a function of the Higgs mass the selection is applied in a mass dependent way.

\section{$W+jet/\gamma$ and Multijet background estimation}

The background contribution from QCD multijet production where jets are misidentified as leptons is estimated from the data by using like-sign lepton events of each analysis which were selected by inverting calorimeter isolation criteria\cite{hohlfeld}. The samples are normalized to data as function of the lepton $p_T^i ~\ (i=1,2)$ in a region of phase space which is dominated by multijet production. In the $\mu + \tau_{had}$ analysis the shape of the $W+jet/\gamma$ background is taken from MC, the normalization however is estimated using data.

\section{Multivariate Techniques}
\begin{figure}[htb]
  \setlength{\unitlength}{1.0cm}
  \begin{center}
    \begin{picture}(12.0,6.2)
      \put(-2.0,0.0){\epsfig{file=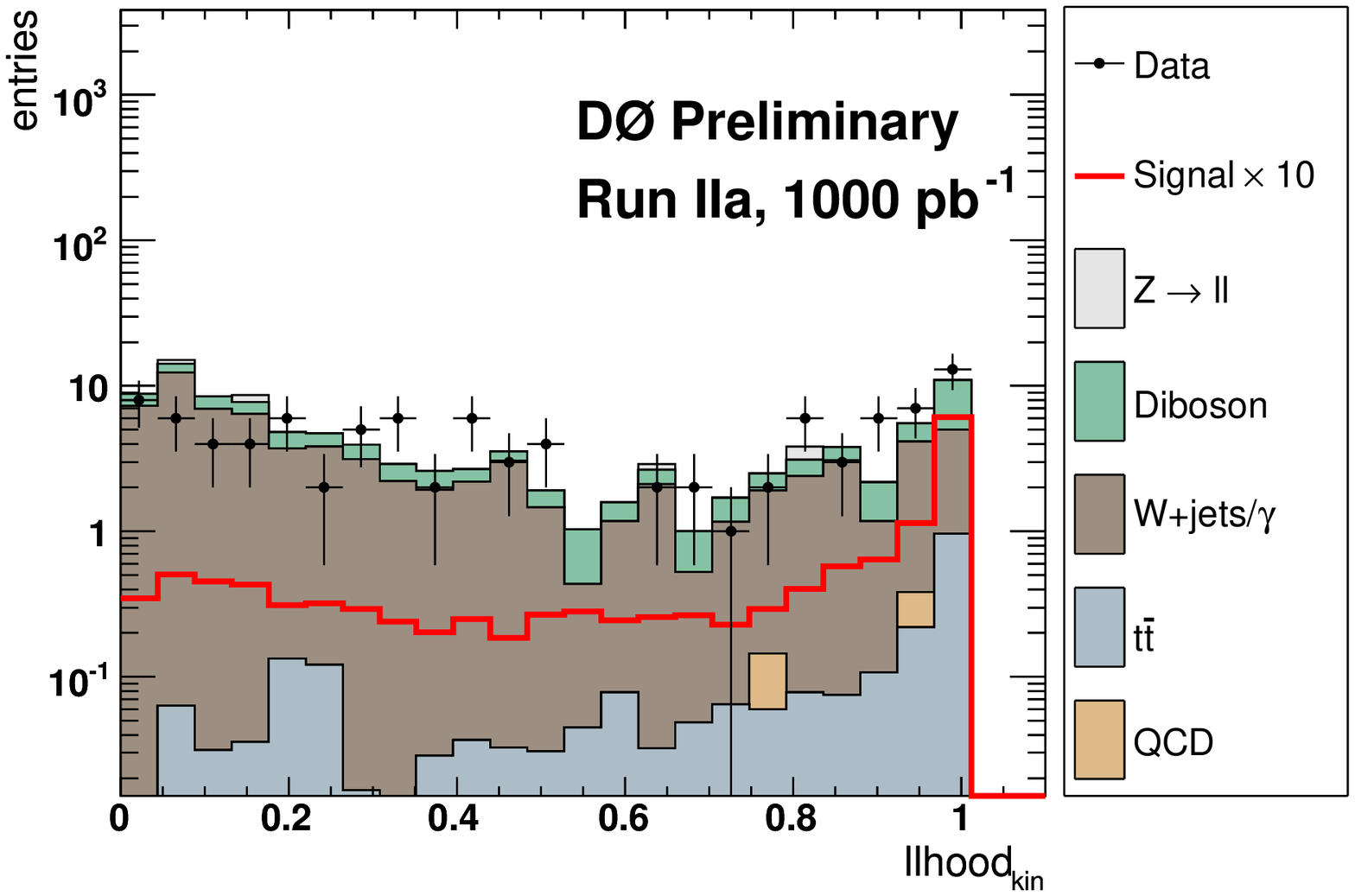, height=6.5cm,width=8cm}}
      \put(6.5,0.0){\epsfig{file=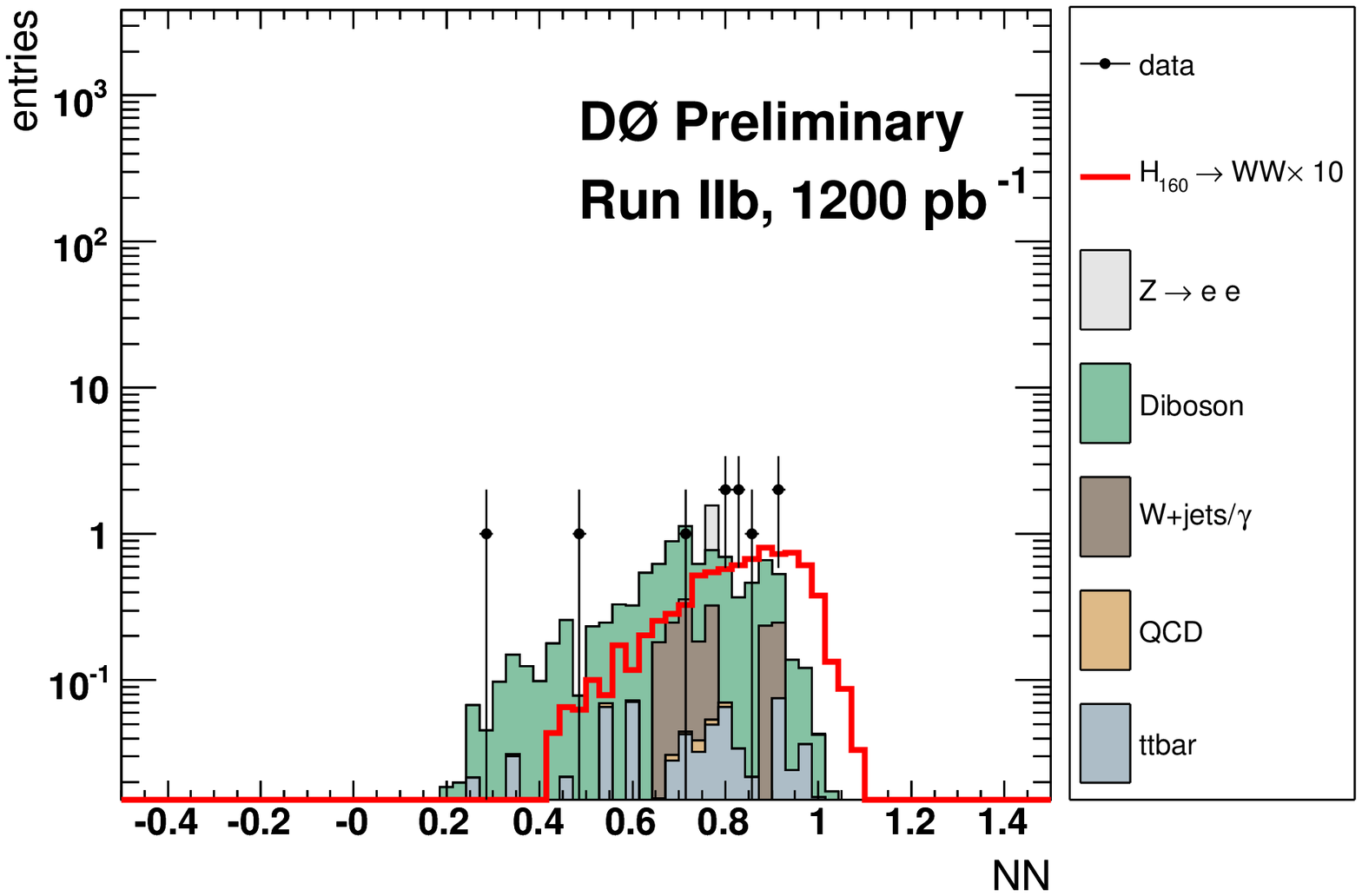, height=6.5cm,width=8cm}}
      \put(-2.2,6.0){a)}
      \put(6.3,6.0){b)}
    \end{picture}
    \caption{\footnotesize
      Distribution of 
      (a) the kinematical likelihood distribution after all selection requirements
      (b) the neural network ouput after all selection requirements for data (points with error bars), background simulation (histograms, complemented with the QCD expectation) and signal expectation for $m_H=160$ GeV (empty histogram). The various signal contributions $H\rightarrow WW \rightarrow ee$ and 
      $H\rightarrow WW \rightarrow \mu\tau_{had}$ are given by the solid red line.
      \label{f:distr1}}
  \end{center}
\end{figure}

At the final stage of the selection as described in Section \ref{ref:sel} the remaining background is dominated by electroweak $W+jets/\gamma$ and diboson production. To improve the background reduction further multivariate techniques have been used, in the $\mu \tau_{had}$-analysis a likelihood approach and for the $ee$ final state artificial neural networks. For the likelihood approach two different likelihoods sensitive to different event properties are used. One likelihood is based on input distributions associated with the selected tau and the second one on kinematical properties of the particular event. All likelihoods are constructed according to formula (\ref{eq_llhood}). A non-negligible fraction of the tau-candidates are electrons misreconstructed as taus which is taken into account by further constructing both classes of likelihoods for that particular events.
\begin{eqnarray}
  {\cal L} =  \frac{{\cal P}_{Sig}(x_1,x_2,\ldots)}{{\cal P}_{Sig}(x_1,x_2,\ldots)+{\cal P}_{Bkgd}(x_1,x_2,\ldots}
  \approx \frac{\prod_i {\cal P}_{Sig}^i}{\prod_i {\cal P}_{Sig}^i+\prod_i {\cal P}_{Bkgd}^i} 
  = \frac{\prod_i {\cal P}_{Sig}^i/{\cal P}_{Bkgd}^i}{\prod_i {\cal P}_{Sig}^i/{\cal P}_{Bkgd}^i+1}
  \label{eq_llhood}
\end{eqnarray} 
Where ${\cal P}_{Sig}^i$ represents the signal and ${\cal P}_{Bkgd}^i$ the background value for a given bin $i$. The value of the input distributions for bin $i$ are given by the variables $x_i$. ${\cal P}_{Sig}^i \equiv {\cal P}_{Sig}(x_i)$ and ${\cal P}_{Sig}^i \equiv {\cal P}_{Sig}(x_i)$ represents the probability density functions for the topological variables. These likelihoods are constructed for each Higgs boson mass point and both tau-types. The resulting likelihood distribution for $m_{\rm H}=160$ GeV and tau-type I is displayed in Fig. \ref{f:distr1}. Using both likelihood classes a further selection requirement is applied. These selections have been optimized for each sample, tau type and Higgs mass. For the dielectron analysis the separation of signal from background is done using an artificial neural network. A separate NN is trained for each Higgs boson mass tested. A list of input variables has been derived based on the separation power of the various distributions. Those variables can be divided into three classes, object kinematics, event kinematics and angular variables. An additional input variable is a discriminant constructed using the matrix element (ME) method. Leading-order parton states for either signal or $WW$ background are integrated over, with each state weighted according to its probability to produce the observed measurement.

\section{Results and Conclusion}
Limits on the cross section for Higgs boson production times the branching fraction into the discussed final states are derived at the 95\% Confidence Level (CL).  Whereas the number of expect signal events, expected background events and observed data is used in the $\mu + \tau_{had}$ analysis, for the dielectron analysis the shape of the NN output distributions are taken into account as well.

\begin{figure}[htb]
  \setlength{\unitlength}{1.0cm}
  \begin{center}
    \begin{picture}(12.0,6.2)
      \put(-2.0,0.0){\epsfig{file=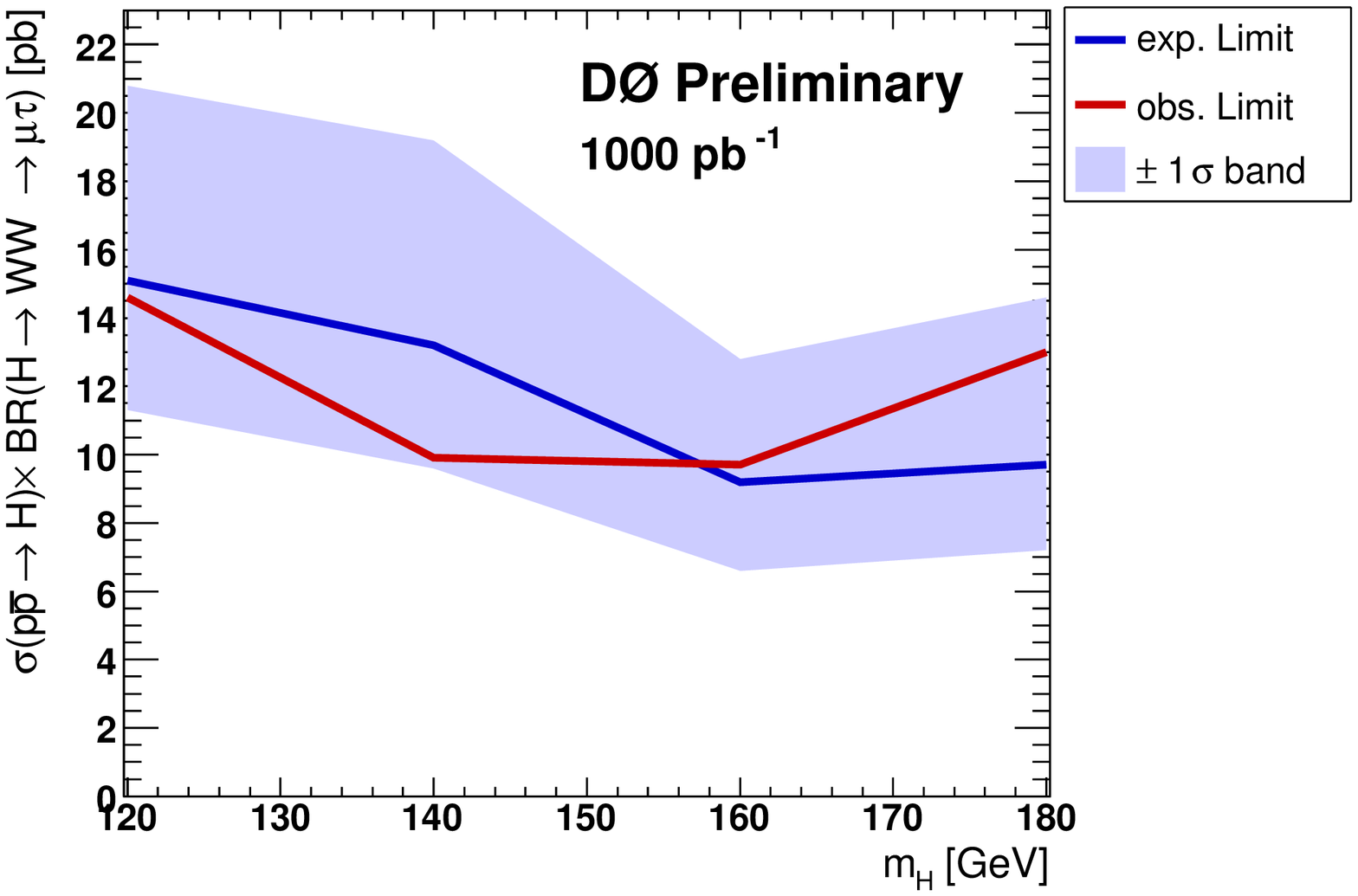, height=6.5cm,width=8cm}}
      \put(6.5,0.0){\epsfig{file=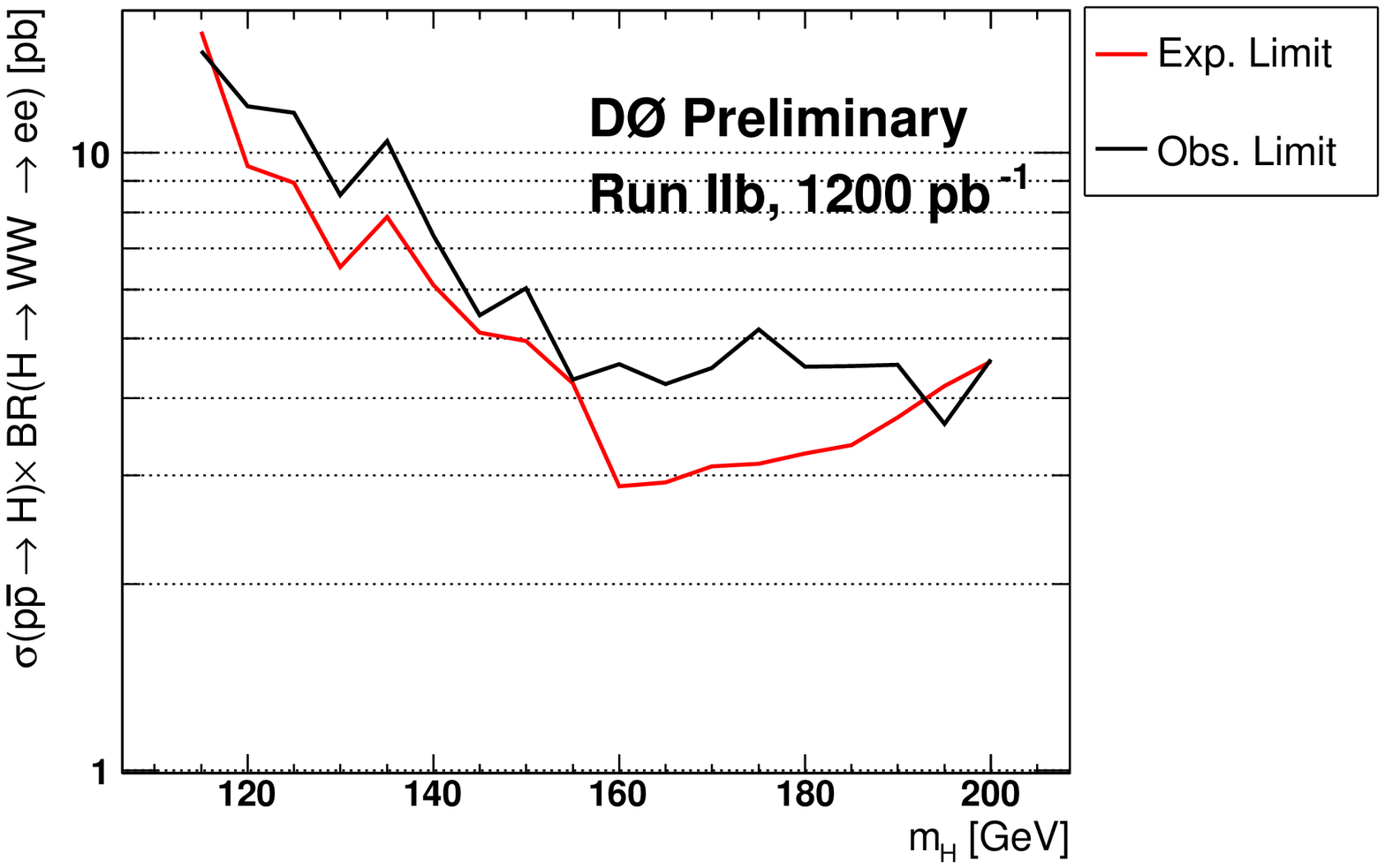, height=6.5cm,width=8cm}}
      \put(-2.2,6.){a)}
      \put(6.3,6.){b)}
    \end{picture}
    \caption{\footnotesize
      (a) Expected and observed limit for the $\mu +\tau_{had}$ analysis
      (b) Expected and observed limit for the dielectron analysis
      \label{f:distr2}}
  \end{center}
\end{figure}

There are various sources of systematic uncertainties included in the calculation of the expected and observed limits:  Lepton identification and reconstruction efficiencies (0.3--13\%), trigger efficiencies (5\%), jet energy scale calibration in signal and background events($<2$\%), track momentum calibration (4\%), detector modeling (1\% for signal, 5-10\% for background), PDF uncertainties ($4$\%), modeling of multijet background (30\%) and theoretical cross section (di-boson 7\%, $t \bar{t}$ 16\%). The uncertainty on the modeling of the electroweak $W+jet/\gamma$ production has been estimated to be 2.5-17.5\%. The systematic uncertainty on the luminosity is mainly a combination of the PDF uncertainty, uncertainty on the NNLO Z cross section (4\%) and data/MC normalization. \\
The total uncertainty on the background is approximately 20\% and for the signal efficiency is 10\%. The effects of these uncertainties on the NN output distribution shapes were also studied and included as additional systematic uncertainties. The expected and observed limits as function of the Higgs mass are shown in Fig. \ref{f:distr2}. After all selection requirements both the neural network output distribution and the likelihood distributions agree with data within their uncertainties with the expected backgrounds. Thus limits are set on the production cross section times branching ratio $\sigma \times BR(H \rightarrow WW^{*})$. We calculate the limits for each channel using a modified frequentist approach  CLs method with a log-likelihood ratio (LLR) test statistic. To minimize the degrading effects of systematics on the search sensitivity, the individual background contributions are fitted to the data observation by maximizing a profile likelihood function for each hypothesis \cite{cite:profiling}. 
\\A search has been performed for the $H\rightarrow WW\rightarrow \ell \ell$ decay signature from the gluon-gluon fusion production of the Standard
 Model Higgs boson in leptonic channels with either  muons and taus or two electrons, using data corresponding to an integrated luminosity altogether of $\approx$ 2.2 fb$^{-1}$.  No evidence for the Higgs particle is observed and no region of the SM Higgs can be excluded. 

\end{document}